# 2D sub-half-wavelength atom localization in a three-level V-type atomic system

Md. Sabir Ali, Vaishali Naik, Ayan Ray, Alok Chakrabarti

Radioactive Ion Beam Facilities Group

Variable Energy Cyclotron Centre, 1/AF Bidhannagar,

Kolkata-700064, India

Email: sabir_vecc@vec.gov.in, sabirbarc@gmail.com

# Abstract

It is shown, using density matrix calculation, that high precision two-dimensional (2D) atom localization in V-type system can be achieved by applying an additional microwave coupling field between the excited states. In the present scheme, two lasers, probe and pump, form a V-type system by coupling the ground state to the excited states while an additional 2D standing microwave field is used to couple the excited states. The solutions of the density matrix equations reveal that the off-diagonal density matrix element $\rho_{31}$, which signify the coherence between the states connected by the probe laser, is position dependent. As the imaginary part of $\rho_{31}$ is proportional to the imaginary part of the susceptibility, the probe absorption also becomes position dependent and in this situation one can readily obtain information about the atom position from the probe absorption spectra. Perfect 2D atom localization can be achieved by carefully selecting different system parameters such as detuning, relative phase, dephasing and field strength. It is found that the atom localization is highly dependent on the probe and coupling field detuning values. In view of this, a systematic study is performed on the system parameters and different zones have been identified which give rise to different absorption shapes, e.g. "dip", "wall", "peak" as well as negative absorption or amplification. Probe absorption spectrum is also found to be dependent on relative phase between the applied fields. The dependence of the dephasing parameter on the probe absorption is investigated and found to be detrimental to atom localization. The effect of the microwave coupling strength on "wall" type absorption shape is also studied and the results are discussed. Finally, a possible practical realization of the present scheme is discussed in Conclusion section.

# I. Introduction

Measuring atomic position with higher precision has always been a fascination for the research community. The appeal lies in surpassing the fundamental limitation imposed by the Heisenberg's uncertainty relation [1] and Rayleigh limit of resolution [2]. The intricate relationship between these two limitations was explained through the thought experiment of Heisenberg's microscope [3] and it seemed that one could not surpass Raleigh limit without violating the Uncertainty relation. Later, in connection with near-field imaging, Vigoureux and Courjon [4] showed that it is possible to go beyond Rayleigh limit without violating the uncertainty relation and the limit is in fact just one point on the uncertainty hyperbola. This was followed by the development of many classical [5,6] as well as quantum techniques [7,8,9] to obtain super-resolution and the general principle of all the techniques have always been to use a phenomenon which is not diffraction limited [10]. These new quantum techniques opened up several new fields such as atom lithography [9,11,12], quantum metrology [13], atom trapping [14], measurement of centre of mass wave function [15,16,17], sub-wavelength microscopy and imaging [18] etc.

In one of the very early experiments [19], the group of Cohen-Tannoudji studied a variation of the Heisenberg's microscope using atomic beam and one-dimensional standing wave. This led Thomas to derive the quantum theory of ultra-high resolution position measurement [20,21]. He and other co-workers also practically demonstrated the technique of position measurement of atoms passing through standing wave optical field [22,23]. In later years, several other works were published aiming at determining the atom position via optical technique [24,25,26,27,28,29,30]. These early methods of atom position measurement are in-depth reviewed by Thomas and Zhang and can be found in ref. [31].

For the last two decades, many other schemes are being introduced regularly for atom localization. One such scheme is to detect spontaneously emitted photons arising from the interaction of moving atoms with a classical standing wave field [**32,33,34**]. Due to the difficult nature of experimental detection of spontaneous emission, another scheme based on the measurement of upper level atomic population [**35,8**] was introduced which led to the subwavelength atom localization. Many other schemes based on the atomic coherence effects, for example, using coherent population trapping [**8**], electromagnetically induced transparency [**17**], double dark resonance [**36**] etc are also very popular in this context. Among other techniques, atom localization via of Autler-Townes doublet [**37**], spontaneously generated coherence [**38**], two photon spontaneous emission detection [**39**] and super-fluorescence [**40**] are few to name.

A new domain of atom localization, termed as sub-half-wavelength localization, was first proposed by Sahrai et al [**41**] in the year 2005 for a four level $\Lambda$ type atomic system. The scheme offers more control and better resolution as the atoms can be localized within a half-wavelength region with this scheme. The experimental detection of the localization is also easy as the scheme makes use of the probe absorption spectrum [**42,41,43**] for localization. In the present work, we investigate one such sub-half-wavelength atom localization in two dimensions (2D) in a V-type system. Similar work for V-type system in one dimension (1D) has been reported by Sahrai and Tajalli [**44**] where they studied the phase dependence of the localization scheme using density matrix calculation. Other reports on V-type atomic system showed the effect of spontaneously generated coherence and detuning on 2D atom localization [**45**], 3D atom localization via three wave mixing [**46**] and atom localization using squeezed vacuum [**47**]. To our knowledge, though 2D atom localization using microwave field has been reported for four level system [48], no has yet reported 2D atom localization in three-level V-type system using microwave coupling field between the upper two levels.

In the present scheme, the probe field connects the lower level with one upper level while the coupling field connects the other upper level to the ground level. A standing microwave coupling field is introduced between the upper two levels which are well separated in energy. This forms a closed cycle transition and the resultant phase dependency was studied by Sahrai and Tajalli [44] for a 1D V-type system. Apart from discussing the phase dependence in the 2D system, it is shown in the results section that the atom localization is highly sensitive to the field detuning values and gives rise to different absorption shapes in different parameter zones. The effect of the dephasing parameter on the 2D localization is explored and is found to be detrimental to the localization. The effect of microwave coupling strength is also studied and discussed at the same time.

## II. Model and equations

The proposed scheme for the atom localization in three-level V-type system is shown in Fig. 1. State $|1\rangle$ represents the ground state while the other two states, $|2\rangle$ and $|3\rangle$, represent the excited states. The laser field connecting the states $|1\rangle$ and $|3\rangle$ is termed as the probe laser whereas the laser which connects the states $|1\rangle$ and $|2\rangle$ is termed as the pump or coupling laser. The presence of an additional two dimensional standing microwave field between the excited states paves the way for 2D atom localization in V-type system. The probe field frequency and phase are denoted as $\omega_p$ and $\phi_p$ respectively whereas the coupling field frequency and phase are $\omega_c$ and $\phi_c$ respectively. The frequency of the microwave field is $\omega_m$ and phase for the same is represented by $\phi_m$. The transition between the states $|2\rangle$ and $|3\rangle$ is electric dipole forbidden and we need microwave field for the same to have non-zero magnetic transition dipole moment between the two states. The coherence decay rates of the states $|2\rangle$ and $|3\rangle$ are identified as $\gamma_2$ and $\gamma_3$ respectively and originate from the decay of population from the excited states to the ground state $|1\rangle$. Transfer of population through decay from state $|3\rangle$ to state $|2\rangle$ is neglected.

The atoms moving along Z direction interacts with a standing microwave field in the *X-Y* plane and it results in position dependent laser atom interaction. The atom velocity in the Z direction is large enough so that spatial interaction in the X-Y plane doesn't alter the motion in Z direction. This assumption enables us to treat the atomic motion in the Z direction classically. We also assume that the centre of mass of the atom is at rest in the interaction plane. Therefore we can neglect the kinetic energy term in the system Hamiltonian following the Raman-Nath approximation [**49,34,50**]. Therefore, under dipole approximation, the total Hamiltonian for the atom-field combined system is given by,

$$H = H_0 + H_I \tag{1}$$

where $H_0$ is the bare atomic Hamiltonian and $H_I$ is the interaction Hamiltonian. The two Hamiltonians can be written as,

$$H_0 = \hbar\omega_1|1\rangle\langle 1| + \hbar\omega_2|2\rangle\langle 2| + \hbar\omega_3|3\rangle\langle 3| \tag{2}$$

and

$$H_I = -[\hbar\Omega_c e^{-i(\omega_c t+\phi_c)}|1\rangle\langle 2| + \hbar\Omega_p e^{-i(\omega_p t+\phi_p)}|1\rangle\langle 3| + \hbar\Omega_m e^{-i(\omega_m t)}|2\rangle\langle 3|] + h.c. \tag{3}$$

$\Omega_c = \vec{E}_c * \vec{\mu}_{12}/2\hbar$ and $\Omega_p = \vec{E}_p * \vec{\mu}_{13}/2\hbar$ are the Rabi frequencies for the transitions $|1\rangle \to |3\rangle$ and $|1\rangle \to |2\rangle$ respectively. $\hbar\omega_i$ $(i = 1, 2, 3)$ is the energy of the level $|i\rangle$. $\vec{E}_p$ and $\vec{E}_c$ are the electric field amplitudes for the probe and coupling field respectively. $\vec{\mu}_{13}$ and $\vec{\mu}_{12}$ are the induced dipole moment of the designated transitions. Equations (2) and (3) are similar to the equations (2-3) in ref. [**44**] with the exceptions that $\Omega_m(x, y) = \Omega_m(Sin\ Kx + Sin\ Ky)$ now represents the two dimensional position dependent interaction between the atom and the field in the form of two dimensional classical standing wave field with the wave vector $K = 2\pi/\lambda$. The total system Hamiltonian under rotating wave approximation can be written as,

$$-\hbar \begin{pmatrix} 0 & e^{i\phi_c}\Omega_c & e^{i\phi_p}\Omega_p \\ e^{-i\phi_c}\Omega_c & \Delta_c & e^{i\phi_m}\Omega_m \\ e^{-i\phi_p}\Omega_p & e^{-i\phi_m}\Omega_m & \Delta_p \end{pmatrix} \quad (4)$$

Here, $\Delta_m = \omega_m - \omega_{23}$, $\Delta_c = \omega_c - \omega_{12}$ and $\Delta_p = \omega_p - \omega_{13}$ are the respective field detuning values corresponding to the atomic transitions $|2\rangle \rightarrow |3\rangle$, $|1\rangle \rightarrow |2\rangle$ and $|1\rangle \rightarrow |3\rangle$. The field detuning values also satisfy the three-photon resonance condition, $\Delta_m = \Delta_p - \Delta_c$. $\omega_{ij}$ ($i, j = 1, 2, 3, i \neq j$) is the angular frequency corresponding to the energy gap between the atomic energy levels. In present calculation, we are interested in finding the atom position information from the probe field absorption. The probe field absorption can be estimated from the electric susceptibility which is directly proportional to the off-diagonal density matrix element $\rho_{31}$ and is given by [**51,52**],

$$\chi = \frac{N|\mu_{13}|^2}{\epsilon_0 \hbar \Omega_p} \rho_{31} \quad (5)$$

here $N$ is the atom number density, $\epsilon_0$ is the permittivity of free space. The susceptibility ($\chi$) can be written as, $\chi = \chi' + i\chi''$ where the probe absorption is proportional to the imaginary part of the susceptibility ($\chi''$) which itself is proportional to $Im(\rho_{31})$. Therefore in the present endeavor, our aim is to calculate steady state values of $Im(\rho_{31})$ for different parameter settings and to understand the behavior of atom localization from the same.

In order to understand the system dynamics we make use of Von-Neumann equation [**53**] for density matrix which can be written as,

$$\dot{\rho} = -\frac{i}{\hbar}[H, \rho] - \frac{1}{2}\{\Gamma, \rho\} \quad (6)$$

where $\{\Gamma, \rho\} = \Gamma\rho + \rho\Gamma$ and $\Gamma = \langle m|\Gamma|n\rangle = \Gamma_m \delta_{mn}$ ($m, n = 1, 2, 3$). $\Gamma$ is the relaxation matrix which incorporates the decay rate in equation (6) and $\Gamma_m$s are the decay rates of the individual levels. The individual decay rates can be replaced with the coherence decay rates $\gamma_{mn} = (\Gamma_m + \Gamma_n)/2$. With the assumptions that $\Gamma_1 = 0$ and upper levels can only decay to the ground state $|1\rangle$, the coherence decay rates simply become $\gamma_{31} \approx \gamma_3 = \Gamma_3/2$ and $\gamma_{21} \approx \gamma_2 = \Gamma_2/2$. The decay between the upper states has been neglected. With the help of equation (6) and (4) the equations of motion for the density matrix elements under steady state condition ($\frac{\partial \rho}{\partial t} \approx 0$) can be written as,

$$2\gamma_2\rho_{22} + 2\gamma_3\rho_{33} - i\left(e^{-i\phi_c}\Omega_c\rho_{12} + e^{-i\phi_p}\Omega_p\rho_{13} - e^{i\phi_c}\Omega_c\rho_{21} - e^{i\phi_p}\Omega_p\rho_{31}\right) = 0 \quad (7)$$

$$\gamma_2\rho_{12} + i\left(e^{i\phi_c}\Omega_c\rho_{11} + \Delta_c\rho_{12} + e^{-i\phi_m}\Omega_m\rho_{13} - e^{i\phi_c}\Omega_c\rho_{22} - e^{i\phi_b}\Omega_p\rho_{32}\right) = 0 \quad (8)$$

$$\gamma_3\rho_{13} + i\left(e^{i\phi_p}\Omega_p\rho_{11} + e^{i\phi_m}\Omega_m\rho_{12} + \Delta_p\rho_{13} - e^{i\phi_c}\Omega_c\rho_{23} - e^{i\phi_p}\Omega_p\rho_{33}\right) = 0 \quad (9)$$

$$\gamma_2\rho_{21} + i\left(-e^{-i\phi_c}\Omega_c\rho_{11} - \Delta_c\rho_{21} + e^{-i\phi_c}\Omega_c\rho_{22} + e^{-i\phi_p}\Omega_p\rho_{23} - e^{i\phi_m}\Omega_m\rho_{31}\right) = 0 \quad (10)$$

$$2\gamma_2\rho_{22} + i\left(-e^{-i\phi_c}\Omega_c\rho_{12} + e^{i\phi_c}\Omega_c\rho_{21} + e^{-i\phi_m}\Omega_m\rho_{23} - e^{i\phi_m}\Omega_m\rho_{32}\right) = 0 \quad (11)$$

$$(\gamma_2 + \gamma_3)\rho_{23} + i\left(-e^{-i\phi_c}\Omega_c\rho_{13} + e^{i\phi_p}\Omega_p\rho_{21} + e^{i\phi_m}\Omega_m\rho_{22} - (\Delta_c - \Delta_p)\rho_{23} - e^{i\phi_m}\Omega_m\rho_{33}\right) = 0 \quad (12)$$

$$\gamma_3\rho_{31} + i\left(-e^{-i\phi_p}\Omega_p\rho_{11} - e^{-i\phi_m}\Omega_m\rho_{21} - \Delta_p\rho_{31} + e^{-i\phi_c}\Omega_c\rho_{32} + e^{-i\phi_p}\Omega_p\rho_{33}\right) = 0 \quad (13)$$

$$(\gamma_2 + \gamma_3)\rho_{32} + i\left(-e^{-i\phi_p}\Omega_p\rho_{12} - e^{-i\phi_m}\Omega_m\rho_{22} + e^{i\phi_c}\Omega_c\rho_{31} + (\Delta_c - \Delta_p)\rho_{32} + e^{-i\phi_m}\Omega_m\rho_{33}\right) = 0 \quad (14)$$

$$2\gamma_3\rho_{33} + i\left(-e^{-i\phi_p}\Omega_p\rho_{13} - e^{-i\phi_m}\Omega_m\rho_{23} + e^{i\phi_p}\Omega_p\rho_{31} + e^{i\phi_m}\Omega_m\rho_{32}\right) = 0 \quad (15)$$

$$\rho_{11} + \rho_{22} + \rho_{33} = 1 \quad (16)$$

Equation (16) follows from the fact that total probability must be equal to unity.

Analytical solution for the equations (7)-(16) can be obtained by assuming $\rho_{11} \cong 1, \rho_{22} \cong \rho_{33} \cong 0$ and $\rho_{23} \cong \rho_{32} \cong 0$. This has been done in ref [**44**] and an expression for $\rho_{31}$ was obtained for one dimensional atom localization on V-type system. Though these are valid assumptions for weak probe field and has been used many times by other researchers [**52,51,48**], we avoided making such assumptions and tried to solve the equations using the computer code *Mathematica*. This enabled us not only to systematically study the dependence of the atom localization on different system parameters without extra assumptions but also new features emerge in the probe absorption line shape with this approach. Using the same approach one can readily obtain an analytical expression for $\rho_{31}$ but it has been avoided to present it in the article due to its bulky appearance. Though it is not readily seen from equations (7)-(16), the probe absorption indeed becomes position dependent in a V-type system and the atom position information can be extracted from the probe absorption as it passes through the standing wave field. Following section discusses the outcome of the calculation in more detail.

### III. Numerical results and discussion

In this section, the results of the calculation on 2D atom localization in V-type system based on the solution of the equations (7)-(16) for $\rho_{31}$ for different values of the system parameters is presented and discussed. The value of the coherence decay rate $\gamma_2$ is assumed to be 1 which is basically same as the normalization of $\rho_{31}$ with respect to $\gamma_2$. The values of the system parameters which have a physical dimension of frequency are in the units of $\gamma_2$. It is also assumed that the phase of the microwave field $\phi_m = 0$ and the relative phase between the coupling and the probe field is represented by $\phi = (\phi_c - \phi_p)$. Following subsections discuss the effects of the system parameters on the atom localization for two distinct cases. In case-I, the detuning values of the probe and the coupling fields are considered to be equal *i.e.* $\Delta_p = \Delta_c$

whereas in case-II they are considered to be unequal i.e. $\Delta_p \neq \Delta_c$. It will be clear from the results that case-II, which was left unexplored by Sahrai and Tajalli [44] in their 1D calculation, exhibits markedly different phenomena than case-I.

### III.A. Case-I: Equal probe and coupling field detuning values ($\Delta_p = \Delta_c$)

In this case, the probe and the coupling field detuning values are assumed to be equal. This situation was explored, under a limited scope, in the earlier work by Sahrai and Tajalli [44] for 1D localization in V-type system. Fig. 2 shows the probe absorption represented by $Im(\rho_{31})$ in $Kx$-$Ky$ plane for two different detuning settings. The locations of the absorption maxima indicate the atom localization region. For $\Delta_p = \Delta_c = -10$, the absorption maxima are located at $x = \lambda/4$, $y = \lambda/4$ and at locations full wavelength apart in both $x$ and $y$ direction as shown in Fig. 2(a). For $\Delta_p = \Delta_c = 10$, the localtions are now shifted at $x = -\lambda/4$, $y = -\lambda/4$ as shown in Fig. 2(b). These two plots show the profound effect of field detuning on the positions of the the localization peaks in $Kx$-$Ky$ plane. In both situations, the obtained maximum values for $Im(\rho_{31})$ are well below 0.4 which signifies that the maximum probability of the localization is less than 50%. The numbers of peaks within a full wavelength in both $X$ and $Y$ direction is also limited to one. On changing the field detuning values from $\Delta_p = \Delta_c = -10$ toward $\Delta_p = \Delta_c = 10$, the appearance of "dip" at the locations of absorption maxima can be seen. This is shown in Fig. 3 (a)-(h). The shape of the dips also changes from circular ones to rectangular ones and back again to the circular ones as the detuning value is changed. The depth of the "dip" also increases and later comes back to the same value as the detuning values are changed. These figures also reveal the gradual shift in the localization positions with the change in the detuning values. While the detuning values are changed, the appearance of a distinctive feature which resembles the appearance of a "wall" type absorption shape can be noticed. Fig. 4 shows this "wall" type absorption shape for near resonant probe and coupling field $(\Delta_p = \Delta_c = -0.3, 0.1)$. In this situation, the probability of absorptions at the earlier atom localization positions $(x, y = \pm \lambda/4)$ are almost close to zero while maximum absorption appears at the "wall" which are located on $Kx + (-1)^l Ky = \pm l\pi$ straight lines on the $Kx$-$Ky$ plane. The absorption peak is much reduced and becomes close to zero for other field detuning values which are not shown here. Next, the behavior of the relative phase ($\phi$) on the probe absorption is studied. Fig. 5 illustrates such phase dependence of the absorption shapes. As $\phi$ is changed from $0$ to $\pi$, the localization positions shift from $x, y = \lambda/4$ to $x, y = -\lambda/4$ along with other localizations peaks at full wavelengths apart. The trend is repeated for the "dip" type absorption shape as well. Though the same phenomenon was reported for 1D localization by Sahrai and Tajalli [**44**], no effect of the phase on the "wall" type absorption shape is noticed for case-I. From this observation one can conclude that profound phase dependency is present at all field detuning values except for near resonant situation even for the 2D case. The effect of the dephasing parameter ($\gamma_3$) on different absorption shapes is presented in Fig. 6. It shows different absorption shapes for two different dephasing parameter values. As $\gamma_3$ is increased from 0.5 to 2.0, the absorption peaks are much reduced which is shown in Fig. 6 (a)-(b). The same is true for "dip" and "wall" type absorption shapes. Therefore it is evident that the dephasing of the excited state affects the atom localization negatively. Lower the dephasing, stronger will be the atom localization. In the final plot for case-I, the effect of the microwave coupling strength($\Omega_m$) on the "wall" type absorption shape is studied. Fig. 7 (a)-(c) show the behavior as $\Omega_m$ is increased from 2.5 to 10. It is clear from the plots that the width of the "wall" is reduced as the coupling strength is increased which means that the absorption line-width on the straight line $Kx + (-1)^l Ky = \pm l\pi$ are affected by the microwave coupling strength adversely.

### III.B. Case-II: Coupling and probe field detuning values are unequal ($\Delta_p \neq \Delta_c$)

In case-II, the nature of the atom localization is studied assuming probe and coupling field detuning values are unequal. This case is pretty different than case-I and brings out some interesting observations. Fig. 8 presents the absorption peaks in *Kx-Ky* space for case-II. Plots (a) to (e) show the absorption peaks for different field detuning values. The number of localization peaks in a full wavelength domain is now doubled compared to case-I. There are now absorption peaks at $x = y = \lambda/4$ and at $x = y = -\lambda/4$ simultaneously. It can also be seen from the plots that the peaks are not of equal height *i.e.* the probability of the localization are not same at the two locations. As the field detuning values are changed, it can be noticed that the localization peak at $-\lambda/4$ diminishes at a faster rate than the localization peak at $\lambda/4$ and ultimately the peak at $-\lambda/4$ becomes stronger compared to the peak at $\lambda/4$. In case-I, as the field detuning values were changed, a shift in the localization position from $-\lambda/4$ to $\lambda/4$ and vice versa was observed but here an alteration in the absorption strength of the localization peaks can be seen. These phenomena for case-II can be considered more general compared to case-I and will be discussed later with the help of Fig. 14. The appearance of negative absorption, as shown in Fig. 8 (f), at $x = y = \lambda/4$ (and at locations full wavelength apart) is also observed in this case. Therefore there is a possibility that the atom localization and light amplification can both be observed under same setting, in this case, the V-type system under the influence of microwave standing wave field. The appearance of "dip" type absorption shape at the locations of the maximum absorption is also observed in case-II for particular field detuning values. The "dip" appears in all of the absorption peaks and it is found that the characteristics of the dips are sensitive to the field detuning values. The diameter and the depth of the dips increased as the probe field detuning values were increased while the coupling field detuning values were kept fixed. This is shown in Fig. 9. The "dip" conveniently converts to "wall" type absorption shape as we keep on increasing the probe field detuning values. Fig. 10 shows the evolution of the "wall" type structure from the "dip" type absorption shape. This conversion takes place within a very narrow range of the probe detuning values $(\Delta_c = -25, \Delta_p = -0.7 \leftrightarrow 0)$ while the other parameters were kept fixed. This overall behavior of the "dip" and the "wall" type absorption shape is very similar the case-I. The effect of the relative phase on localization for case-II is studied next. It is found that the effect is somewhat similar to case-I and is shown in Fig. 11. The absorption or localization "peaks" do change their locations from $x, y = \lambda/4$ to $x, y = -\lambda/4$ and vice versa as the phase $\phi$ is changed from 0 to $\pi$. This also holds true for other locations at full wavelength apart. As the absorption peaks at these locations are not of equal height, this change is readily observable and is shown in Fig. 11(a)-(b). The same is true for the "dip" type absorption shapes and is illustrated in Fig. 11(c)-(d). Plot (e) and (f) show the response of the "wall" type absorption shapes under the influence of $\phi$ and no noticeable change in its appearance can be found. The next job was to monitor the effect of $\gamma_3$ on the probe absorption and similar behavior as in case-I is seen for all the absorption shapes. This is presented in Fig. 12 (a)-(f). In a final attempt to identify any other marked difference between case-I and case-II, the effect of $\Omega_m$ on the "wall" type absorption shape was also studied for case-II but no noticeable difference can be pointed out. Though the absorption line-width is not quantified in this attempt, it can be seen from Fig. 13 that the "wall" thickness is reduced as we increased the microwave coupling strength $(\Omega_m)$.

From Case-I and Case-II it is pretty clear that the absorption shapes are very sensitive to the field detuning and a little change in their value can give rise to different types of absorption shapes. In view of this, it is desirable to have better insight in the system parameters so that different absorption shapes can be produced predictably just by tweaking the system parameters. This led us to Fig. 14 where probe absorption at $x, y = \pm \lambda/4$ is plotted for different $(\Delta_c, \Delta_p)$ combination and for different values of $\Omega_m$. In Fig. 14 (i) and (ii), the behavior of the probe absorption at $\lambda/4$ is shown for $\phi = 0$ and $\pi$ respectively. The Z-axis of the plot represents $Im[\rho_{31}(\Omega_m = 1) + \rho_{31}(\Omega_m = 5) + \rho_{31}(\Omega_m = 10)]$. Separate hill and trenches can be seen on the $\Delta_c$-$\Delta_p$ plane which corresponds to these three different values of $\Omega_m$. The field detuning values at the peak positions on the hills correspond to the appearance of nice looking localization peaks on the *Kx-Ky* plane whereas the bottom of the trenches corresponds to the appearance of the negative absorption peaks or light amplification. The detuning values on the convex side surface of the hill correspond to the appearance of "dip" and "wall" type absorption shapes on the *Kx-Ky* plane. For the detuning values on the concave side surface of the hill, the height of the absorption peak is gradually reduced while moving down the hill and eventually it becomes close to zero but no other absorption shapes can be seen on *Kx-Ky* plane. It is also important to note that there is no absorption on the second and fourth quadrant on the $\Delta_c$-$\Delta_p$ plane. In Fig. 14 (iii), a contour plot is also shown for $\Omega_m = 10$. The behavior of the probe absorption at $-\lambda/4$ is also investigated. The result looks similar to Fig. 14 (ii) and is presented in Fig. 14 (iv). These plots altogether can give an insight to the $(\Delta_c, \Delta_p)$ combination to choose from which can serve the purpose. Finally while comparing Fig. 14 (i) with Fig. 14 (iv), just by drawing an imaginary straight line which satisfy $\Delta_c = \Delta_p$ on the $\Delta_c$-$\Delta_p$ plane, one can readily see that for any particular $\Omega_m$ if there is an absorption peak at $\lambda/4$ site, the absorption is zero at $-\lambda/4$ and vice versa. This symmetry is absent for $\Delta_c \neq \Delta_p$ and simultaneous absorption peaks at both $-\lambda/4$ and $\lambda/4$ locations are present. This may be due to the imposition of additional two-photon resonance condition in case-I on top of the assumed three-photon resonance condition.

The additional two-photon resonance condition can allow for destructive quantum interference to occur at the alternate sites which might lead to missing absorption peaks at the alternate sites. However, one can only tell for sure when better insight into the system dynamics is gained.

## IV. Conclusions

In the present study, the possibility of extending an earlier proposed one dimensional atom localization scheme to two dimension for a V-type three level atomic system is explored. Two orthogonal standing microwave field in *x* and *y* direction is chosen to drive the atomic transition between the upper two atomic levels while the same levels form a conventional V-type coupling to the ground level with the help of a probe field and a coupling field. This is in contrast to the earlier localization scheme where only a single standing-wave field in *x* direction is used between the upper levels. The results show the presence of localization peaks for two dimensional system and therefore it is claimed that the scheme of localization can be suitably extended to two dimension. It is found that the number of localization peak per period of the standing wave field is equal to one for equal probe and coupling field detuning values and it doubles when the detuning values are not equal. The dependence of the atom localization on different field parameters *i.e.* detuning, relative phase, dephasing and microwave coupling strength is also studied. It is observed that relative phase between the probe and the coupling field is a crucial parameter and changing it from 0 to $\pi$ either changes the position of the localization peaks on the localization plane (*Kx-Ky*) or alters the localization probability depending on the field detuning values. The appearance of different other absorption shapes e.g. "dip", "wall" has also been marked on the $\Delta_c$-$\Delta_p$ plane which will certainly help in identifying the optimum condition for achieving the localization. Finally, in ref [**54,55**], the authors used Cesium $6S_{1/2}$ ($F = 4$) energy level as the ground state and $6P_{3/2}$ ($F' = 3,4,5$) as the excited states to form a typical V-type atomic configuration using two lasers. Since the transition frequencies within the hyperfine manifold of $6P_{3/2}$ energy level fall in the microwave region, practical realization of the scheme should be possible by sending Cesium atoms through a two dimensional microwave cavity and using two appropriate lasers.

# References

[1] W. Heisenberg, "Über den anschaulichen Inhalt der quantentheoretischen Kinematik und Mechanik," *Zeitschrift für Physik*, vol. 43, no. 3-4, pp. 172-198, March 1927.

[2] Lord Rayleigh, "On the theory of optical images, with special reference to the microscope," *The London, Edinburgh, and Dublin Philosophical Magazine and Journal of Science*, vol. 42, no. 255, 1896.

[3] Warner Heisenberg, *The physical principles of the quantum theory*. New York: Dover Publications, Inc, 1949.

[4] J. M. Vigoureux and D. Courjon, "Detection of nonradiative fields in light of the Heisenberg uncertainty principle and the Rayleigh criterion," *Applied Optics*, vol. 31, no. 16, pp. 3170-3177, 1992.

[5] C. Girard and A. Dereux, "Near-field optics theories," *Reports on Progress in Physics*, vol. 59, pp. 657-699, 1996.

[6] J. B. Pendry, "Negative refraction makes a perfect lens," *Physical Review Letters*, vol. 85, pp. 3966-3969, 2000.

[7] E.A. Paspalakis, F. Terzis, P.L. Knight, "Quantum interference induced sub-wavelength atomic localization," *Journal of Modern Optics*, vol. 52, pp. 1685-1694, 2005.

[8] G.S. Agarwal and K. T. Kapale, "Subwavelength atom localization via coherent population trapping," *Journal of Physics B*, vol. 39, pp. 3437-3446, 2006.

[9] K.S. Johnson et al, "Localization of metastable atom beams with optical standing waves:

Nano-lithography at the heisenberg limit," *Science*, vol. 280, pp. 1583-1586, 1998.

[10] Kishore T. Kapale, "Subwavelength Atom Localization," in *Progress in Optics*. Amsterdam: Elsevier, 2013, pp. Chapter 4, 199250.

[11] A. N. Boto et al, "Quantum interferometric optical lithography: Exploiting entanglement to beat the diffraction limit," *Physical review letters*, vol. 85, pp. 2733-2736, 2000.

[12] L. Jin, H. Sun, Y. Niu, S. Jin, S. Gong, "Two-dimension atom nano-lithograph via atom localization," *Journal of Modern Optics*, vol. 56, pp. 805-810, 2009.

[13] V. Giovannetti, S. Lloyd, and L. Maccone, "Quantum metrology," *Physical Review letters*, vol. 96, no. 010401, pp. 1-4, 2006.

[14] J. Mompart, V. Ahufinger, and G. Birkl, "Coherent patterning of matter waves with subwavelength localization," *Physical Review A*, vol. 79, no. 053638, 2009.

[15] Kishore T. Kapale, Shahid Qamar, M. Suhail Zubairy, "Spectroscopic measurement of an atomic wave function," *Physical Review A*, vol. 67, no. 023805, 2003.

[16] J. Evers, Shahid Qamar, and M. Suhail Zubairy, "Atom localization and center-of-mass wave function determination via multiple simultaneous quadrature measurements," *Physical Review A*, vol. 75, no. 053809, 2007.

[17] N. A. Proite, Z. J. Simmons, and D. D. Yavuz, "Observation of atomic localization using electromagnetically induced transparency," *Physical Review A*, vol. 83, no. 041803(R), 2011.

[18] K. T. Kapale and G. S. Agarwal, "Subnanoscale resolution for microscopy via coherent

population trapping," *Optics Letters*, vol. 35, pp. 2792-2794, 2010.

[19] C. Cohen-Tannoudji et al, "Channeling atoms in a laser standing wave," *Physical Review Letters*, vol. 59, no. 15, p. 1659, October 1987.

[20] J. E. Thomas, "Uncertainity-limited position measurement of moving atoms using optical fields," *Optics Letters*, vol. 14, no. 21, pp. 1186-1188, November 1989.

[21] J. E. Thomas, "Quantum theory of atomic position measurement using optical fields," *Physical Review A*, vol. 42, no. 9, pp. 5652-5666, November 1990.

[22] K. D. Stokes et al, "Precision position measurement of moving atoms using optical fields," *Physical Review Letters*, vol. 67, no. 15, pp. 1997-2000, October 1991.

[23] J. R. Gardner et al, "Suboptical wavelength position measurement of moving atoms using optical fields," *Physical Review Letters*, vol. 70, no. 22, pp. 3404-3407, May 1993.

[24] Pippa Storey et al, "Measurement-induced diffraction and interference of atoms," *Physical Review Letters*, vol. 68, no. 4, pp. 472-475, January 1992.

[25] Stefan Kunze et al, "Diffraction of atoms from a measurement induced grating," *Physical Review letters*, vol. 78, no. 11, pp. 2038-2041, March 1997.

[26] Ralf Quadt, "Measurement of atomic motion in a standing light field by homodyne detection," *Physical Review Letters*, vol. 74, no. 3, pp. 351-354, January 1995.

[27] R. Quadt, M. Collet, D. F. Walls, "Measurement of atomic motion in a standing light field by homodyne detection," *Physical Review Letter*, vol. 74, pp. 351-354, 1995.

[28] F. L. Kien, G. Rempe, W. P. Schleich, M. S. Zubairy, "Atom localization via Ramsey interferometry: a coherent cavity field provides a better resolution," *Physical Review A*, vol. 56, pp. 2972-2977, 1997.

[29] M. A. M. Marte and P. Zoller, "Quantum non demolition measurement of transverse atomic position in Kapitza-Dirac atomic beam scattering," *Applied Physics B: Photophysics and laser chemistry*, vol. 54, pp. 477-485, 1992.

[30] S. Kunze, K. Dieckmann, G. Rempe, "Diffraction of atoms from a measurement induced grating," *Physical Review Letter*, vol. 78, pp. 2038-2041, 1997.

[31] J. E. Thomas and L. J. Wang, "Precision position measurement of moving atoms," *Physics Reports*, vol. 262, pp. 311-366, 1995.

[32] A. M. Herkommer, W. P. Schleich, M. S. Zubairy, "Autler-Townes microscopy on a single atom," *Journal of Modern Optics*, vol. 44, p. 2507, 1997.

[33] M. Holland, S. Marksteiner, P. Marte, P. Zoller, "Measurement induced localization from spontaneous decay," *Physical Review Letter*, vol. 76, pp. 3683-3686, 1996.

[34] Fazal Ghafoor, Sajid Qamar, and M. Suhail Zubairy, "Atom localization via phase and amplitude control of the driving field," *Physical Review A*, vol. 65, no. 043819, pp. 1-8, 2002.

[35] E. Paspalakis and P. L. Knight, "Localizing an atom via quantum interference," *Physical Review A*, vol. 63, no. 056802, 2001.

[36] D.-C. Cheng, Y.-P. Niu, R.-X. Li, S.-Q. Gong, "Controllable atom localization via double

dark resonances in a tripod system," *Journal of the Optical Society of America B*, vol. 23, pp. 2180-2184, 2006.

[37] S. Qamar, S. Y. Zhu, M. S. Zubairy, "Precision localization of single atom using Autler–Townes microscopy," *Optics Communications*, vol. 176, pp. 409-416, 2000.

[38] M. Sahrai, "Atom localization via absorption spectrum," *Laser Physics*, vol. 17, pp. 98-102, 2007.

[39] S. Qamar, J. Evers, M. S. Zubairy, "Atom microscopy via two-photon spontaneous emission spectroscopy," *Physical Review A*, vol. 79, no. 043814, 2009.

[40] M. Macovei, J. Evers, C. H. Keitel, M. S. Zubairy, "Localization of atomic ensembles via super-fluorescence," *Physical Review A*, vol. 75, no. 033801, 2007.

[41] M. Sahrai, H. Tajalli, K. T. Kapale, M. S. Zubairy, "Subwavelength atom localization via amplitude and phase control of the absorption spectrum," *Physical Review A*, vol. 72, no. 013820, 2005.

[42] H. Tajalli, M. Sahrai, "Atom localization via electromagnetically induced transparency," *Laser Physics*, vol. 7, pp. 1007-1011, 2004.

[43] M. Sahrai, M. Mahmoudi, R. Kheradmand, "Atom localization of a two-level pump-probe system via the absorption spectrum," *Laser Physics*, vol. 17, pp. 40-44, 2007.

[44] Mostafa Sahrai and Habib Tajalli, "Sub half-wavelength atom localization of a V-type three-level atom via relative phase," *Journal of Optical Society of America*, vol. 30, no. 3, pp. 512-517, 2013.

[45] Zhao Shun-Cai et al, "Effect of spontaneously generated coherence and detuning on 2D atom localization in two orthogonal standing-wave fields," *Chinese Physics Letters*, vol. 31, no. 034206, pp. 1-5, 2014.

[46] Zhonghu Zhu et al, "High-precision three-dimensional atom localization via three-wave mixing in V-type three-lecvel atoms," *Physics Letters A*, vol. 380, pp. 3956-3961, 2016.

[47] Jun Xu and Fei Wang, "Two-dimensional atom localization induced by a squeezed vacuum," *Chinese Physics B*, vol. 25, no. 104201, pp. 1-10, 2016.

[48] Chunling Ding, Jiahua Li, Xiaoxue Yang, Duo Zhang, and Hao Xiong, "Proposal for efficient two-dimensional atom localization using probe absorption in a microwave-driven four-level atomic system," *Physical Review A*, vol. 84, no. 043840, pp. 1-9, 2011.

[49] Sajid Qamar, Shi-Yao Zhu, and M. Suhail Zubairy, "Atom localizatio via resonance fluorescence," *Physical Review A*, vol. 61, no. 063806, pp. 1-5, 2000.

[50] P. Meystre and M. Sargent, *Elements of quantum optics*. Tucson: Springer, 2007.

[51] Rahmatullah and Sajid Qamar, "Two-dimensiional atom localization via probe-absorption spectrum," *Physical Review A*, vol. 88, no. 013846, pp. 1-7, 2013.

[52] H. R. Hamedi and Gediminas Juzeliunas, "Phase-sensitive atom localization for closed-loop quantum systems," *Physical Review A*, vol. 94, no. 013842, pp. 1-11, 2016.

[53] Marlan O. Scully and M. Suhail Zubairy, *Quantum Optics*.: Cambridge University Press, 1997.

[54] Jianming Zhao et al, "Experimental measurement of absorption and dispersion in V-type

cesium atom," *Optics Communications*, vol. 206, no. 4-6, pp. 341-345, June 2002.

[55] A Ray et al, "Frequency stabilization of a diode laser using electromagnetically induced transparency in a V-configuration Cesium atom," *Laser Physics*, vol. 17, no. 12, pp. 1353-1360, December 2007.

# Figure captions

Fig. 1 Coupling scheme for the three level V type system. $\omega_m$ represents the two dimensional standing microwave field frequency while $\omega_p$ and $\omega_c$ represents the frequencies of the probe and coupling laser respectively. $\Gamma_2$ and $\Gamma_3$ are the decay rates of the excited states.

Fig. 2 Probe absorption presented by $Im(\rho_{31})$ in two-dimensional $Kx$-$Ky$ space indicating atom localization for $\Delta_p = \Delta_c$. Field detuning values are (a) $\Delta_p = \Delta_c = -10$ and (b) $\Delta_p = \Delta_c = 10$. The other parameters are, $\Omega_p = 0.25$, $\Omega_c = 0.25$, $\Omega_m = 5.0$, $\gamma_3 = 0.5$ and $\phi = 0$. All parameters are in the units of $\gamma_2$. These parameters were kept unchanged for other plots unless specified.

Fig. 3 Probe absorptions for different probe and coupling field detuning values for case-I. Dip appears at the center of the absorption peaks and its shape changes with the detuning. The detuning for the respective plots are, (a) $\Delta_p = \Delta_c = -9$ (b) $\Delta_p = \Delta_c = -5$ (c) $\Delta_p = \Delta_c = -1.5$ (d) $\Delta_p = \Delta_c = 0.8$ (e) $\Delta_p = \Delta_c = 1.6$ (f) $\Delta_p = \Delta_c = 3.5$ (g) $\Delta_p = \Delta_c = 7.0$ (h) $\Delta_p = \Delta_c = 9.0$.

Fig. 4 Probe absorptions for different probe and coupling field detuning values for case-I. The earlier dip type structure has been converted to a "wall" type structure. The detuning for the respective plots are, (a) $\Delta_p = \Delta_c = -0.3$ (b) $\Delta_p = \Delta_c = 0.1$.

Fig. 5 Dependence of probe absorption on relative phase ($\phi$) between the probe and coupling field for different field detuning. $\phi = 0$ for (a), (c) and (e) whereas $\phi = \pi$ for (b), (d) and (f). The respective field detuning values are, (a)-(b) $\Delta_p = \Delta_c = -10$ (c)-(d) $\Delta_p = \Delta_c = 10$ and (e)-(f) $\Delta_p = \Delta_c = -9$.

Fig. 6 Dependence of probe absorption on dephasing parameter ($\gamma_3$) of the excited state $|3\rangle$. The effect is shown for three different probe absorption scenarios which are peak, dip and wall for two different values of $\gamma_3$. For (a), (c) and (e) $\gamma_3 = 0.5$ whereas for (b), (d) and (f) $\gamma_3 = 2.0$. The detuning for the respective plots are (a)-(b) $\Delta_p = \Delta_c = -10$, (c)-(d) $\Delta_p = \Delta_c = -9$ and (e)-(f) $\Delta_p = \Delta_c = 0.1$.

Fig. 7 Dependence of "wall thickness" on microwave coupling strength ($\Omega_m$). The chosen microwave coupling strengths are (a)$\Omega_m = 2.5$, (b) $\Omega_m = 5$ and (c) $\Omega_m = 10$. The field detunings are $\Delta_p = \Delta_c = 0$.

Fig. 8 Probe absorption in two dimensional $Kx$-$Ky$ space indicates atom localization for case-II ($\Delta_p \neq \Delta_c$). Used field detuning values for the plots are (a) $\Delta_c = -25, \Delta_p = -4.1$ (b) $\Delta_c = -12, \Delta_p = -8.4$ (c) $\Delta_c = -3, \Delta_p = -34.0$ (d) $\Delta_c = 12, \Delta_p = 8.3$ (e) $\Delta_c = 25, \Delta_p = 4.1$ (f) $\Delta_c = 5, \Delta_p = 21$. Plot (f) shows the simultaneous appearance of negative and positive absorption peaks.

Fig. 9 Probe absorption with a central "dip" for different field detuning settings for case-II. Field detuning values are (a) $\Delta_c = -25, \Delta_p = -3.7$, (b) $\Delta_c = -25, \Delta_p = -3.1$ and (c) $\Delta_c = -25, \Delta_p = -1.6$.

Fig. 10 Probe absorption showing the appearance of "wall" type structure for different field detuning for case-II. The field detuning values for the respective plots are (a) $\Delta_c = -25, \Delta_p = -0.7$ (b) $\Delta_c = -25, \Delta_p = -0.4$ and (c) $\Delta_c = -25, \Delta_p = 0$.

Fig. 11 Dependence of probe absorption on relative phase ($\phi$) between the probe and coupling field for case-II. Different probe and coupling field detuning values give rise to different absorption shapes (*i.e.*"peak", "dip", "wall" etc) in the two-dimensional $Kx$-$Ky$ space. The detuning values for the respective plots are (a)-(b)$\Delta_c = -25, \Delta_p = -4.1$ (c)-(d)$\Delta_c = -25, \Delta_p = -3.1$ and (e)-(f)$\Delta_c = -25, \Delta_p = 0$. $\phi = 0$ for (a), (c) and (e) whereas $\phi = \pi$ for (b), (d) and (f).

Fig. 12 Dependence of probe absorption on dephasing parameter $\gamma_3$ for case-II. The dependence is shown for different absorption shapes. The detuning values for the plots are (a)-(b) $\Delta_c = -25, \Delta_p = -4.1$ (c)-(d) $\Delta_c = -25, \Delta_p = -3.1$ and (e)-(f) $\Delta_c = -25, \Delta_p = 0$. For (a), (c) and (e) $\gamma_3 = 0.5$ whereas for (b), (d) and (f) $\gamma_3 = 2.0$.

Fig. 13 Dependence of "wall thickness" on microwave coupling strength ($\Omega_m$) for case-II. The chosen microwave coupling strengths are (a) $\Omega_m = 2.5$, (b) $\Omega_m = 5$ and (c) $\Omega_m = 10$. The chosen field detuning values are $\Delta_c = -25, \Delta_p = 0$.

Fig. 14 3D Plot of probe absorption in $\Delta_c$-$\Delta_p$ plane for (i) $x = y = \lambda/4, \phi = 0$ (ii) $x = y = \lambda/4, \phi = \pi$ (iv) $x = y = -\lambda/4, \phi = 0$. The microwave coupling strength chosen for plots (i), (ii) and (iv) are (a)-(d) $\Omega_m = 1$, (b)-(e) $\Omega_m = 5$ and (c)-(f) $\Omega_m = 10$. Plot (iii) shows contour-plot of probe absorption for $\Omega_m = 10$ and $\phi = 0$ at $x = y = \lambda/4$.

**Figures:**

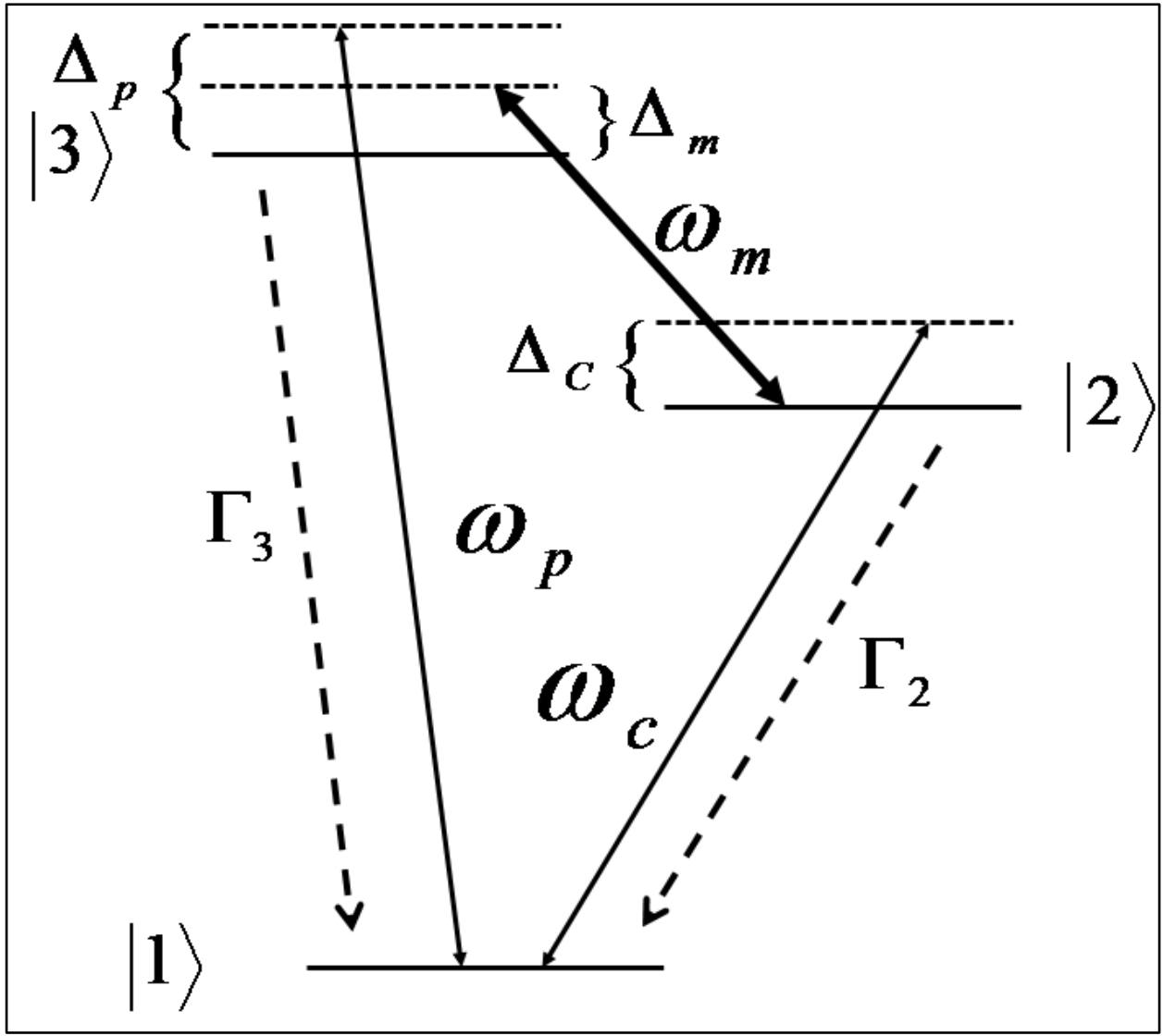

Fig. 1

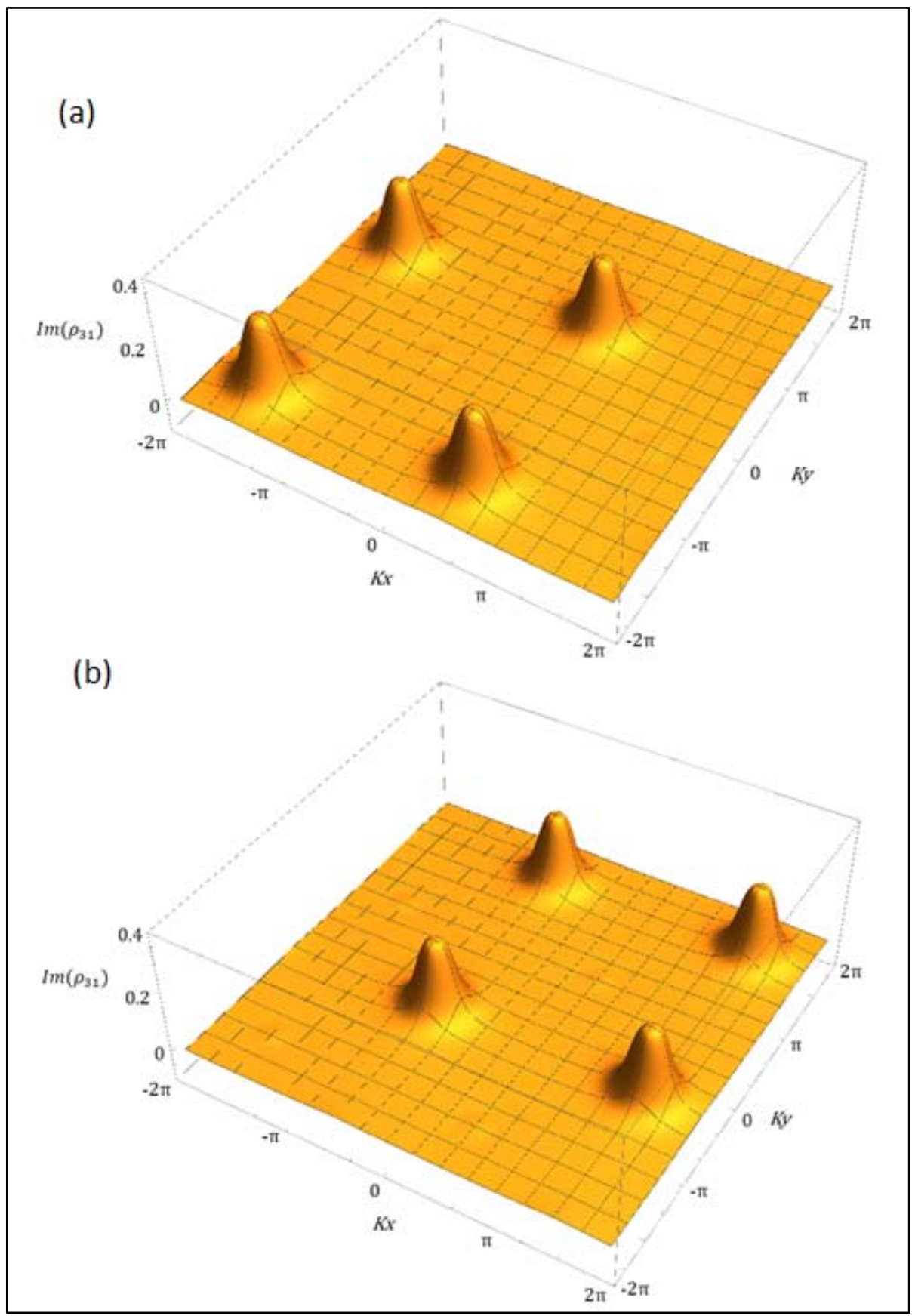

Fig. 2

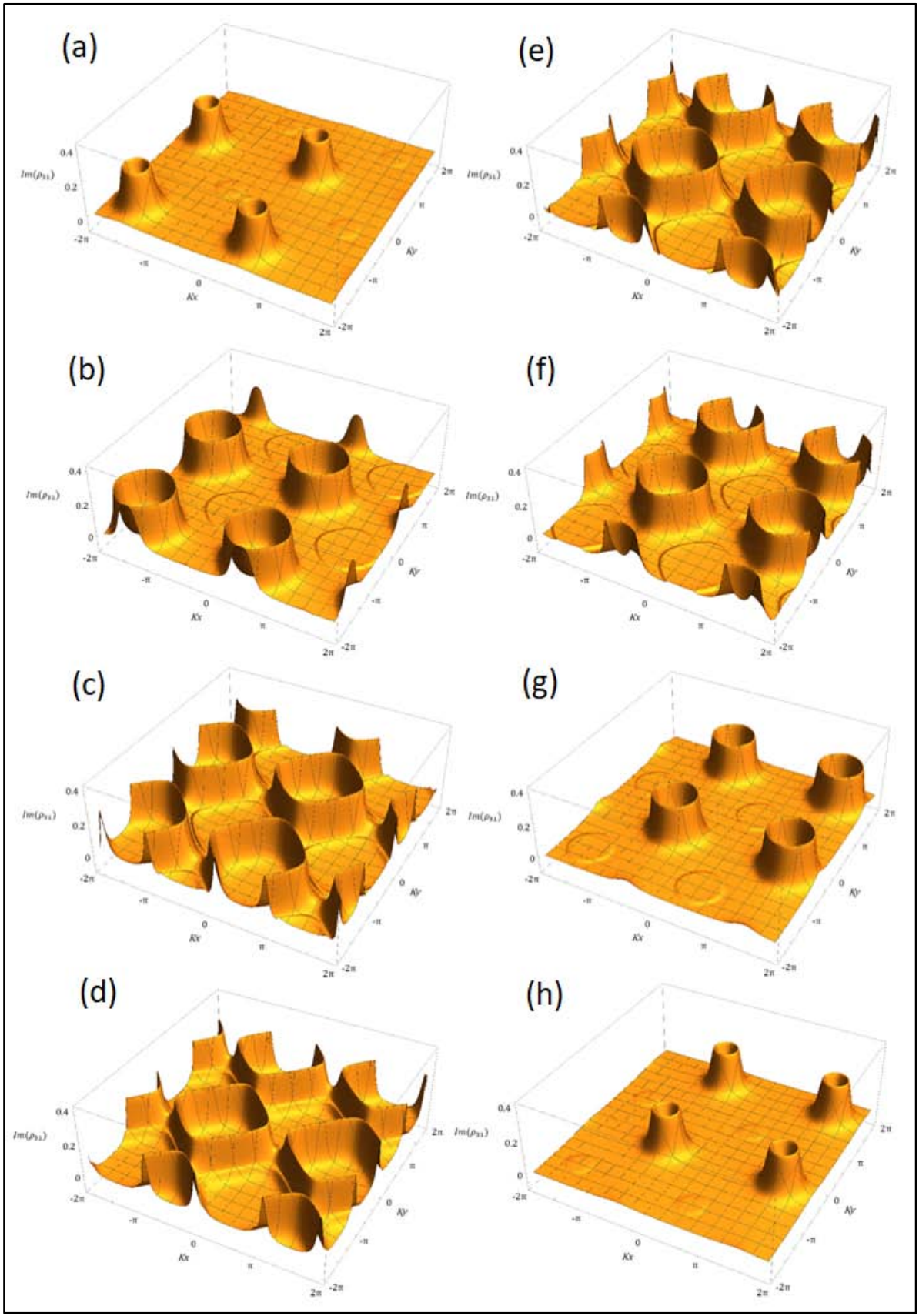

Fig. 3

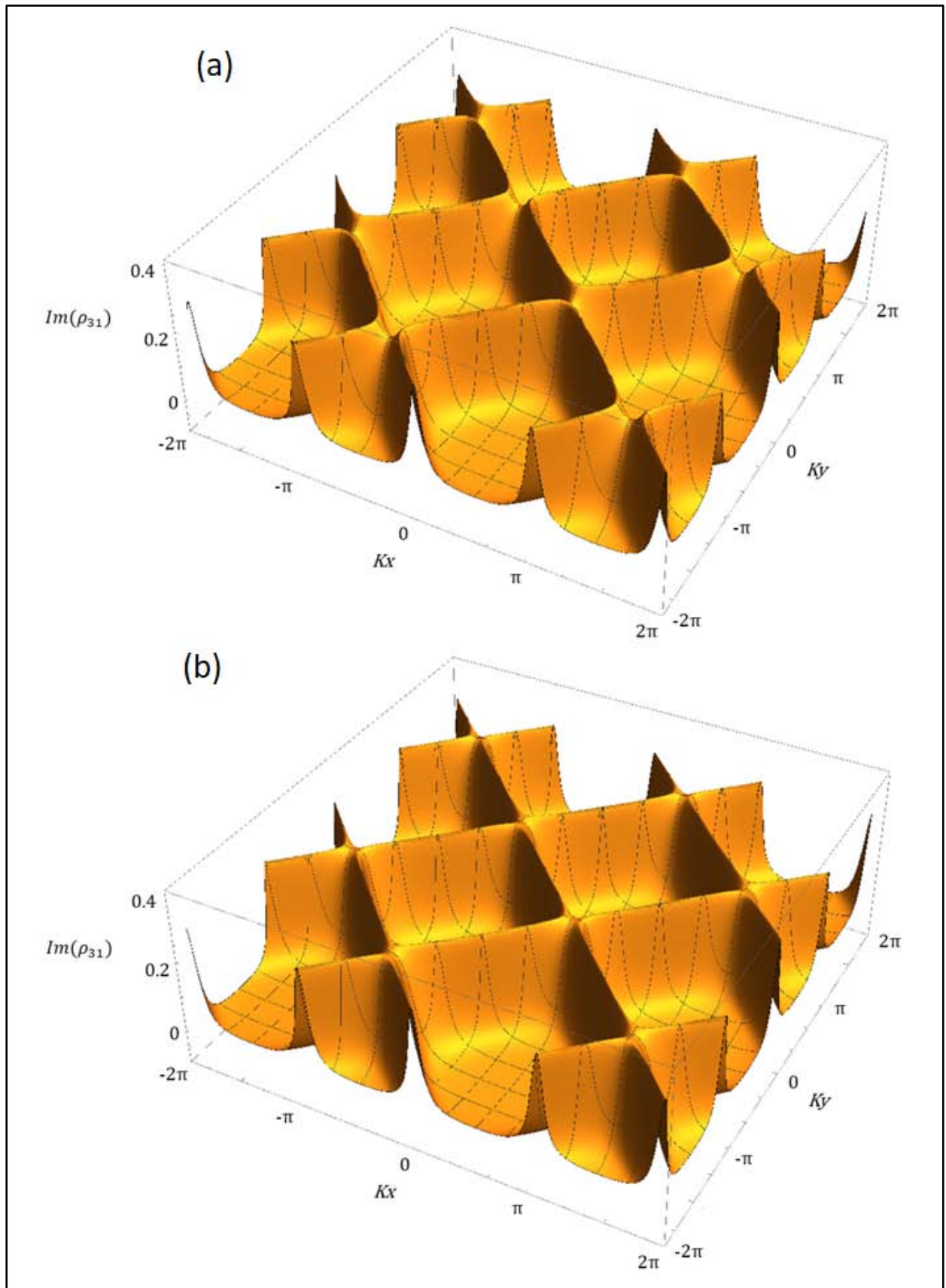

Fig. 4

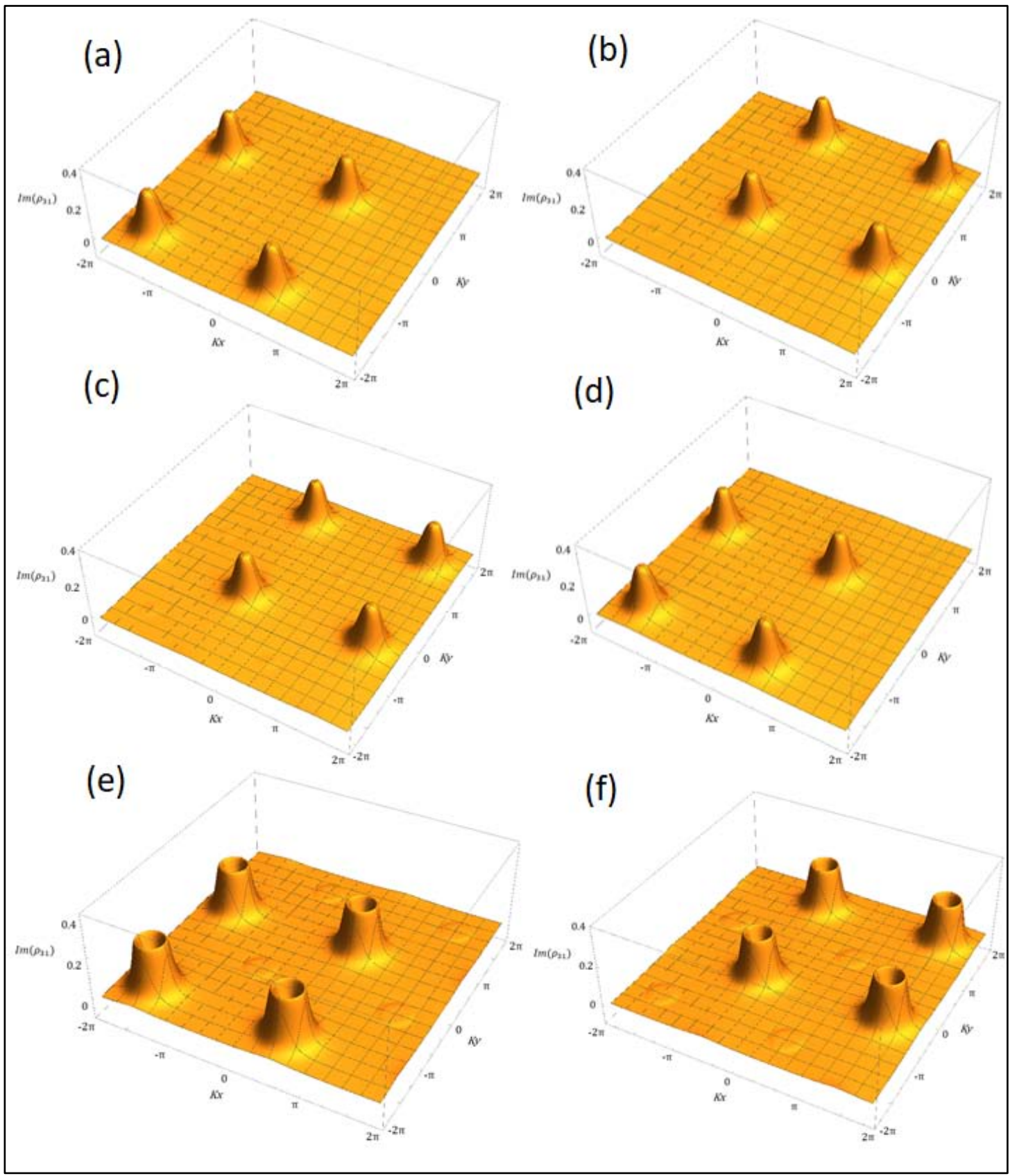

Fig. 5

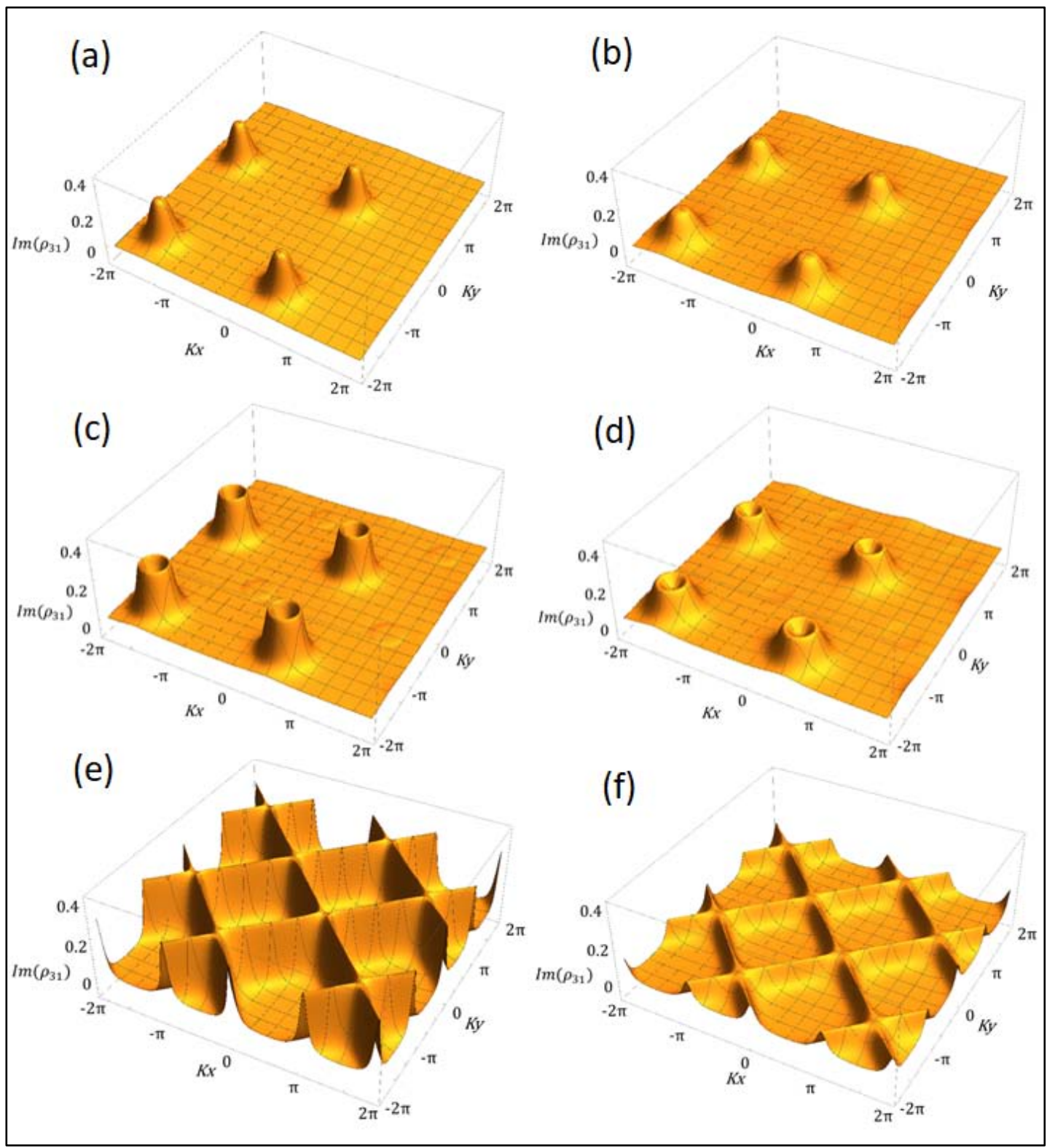

Fig. 6

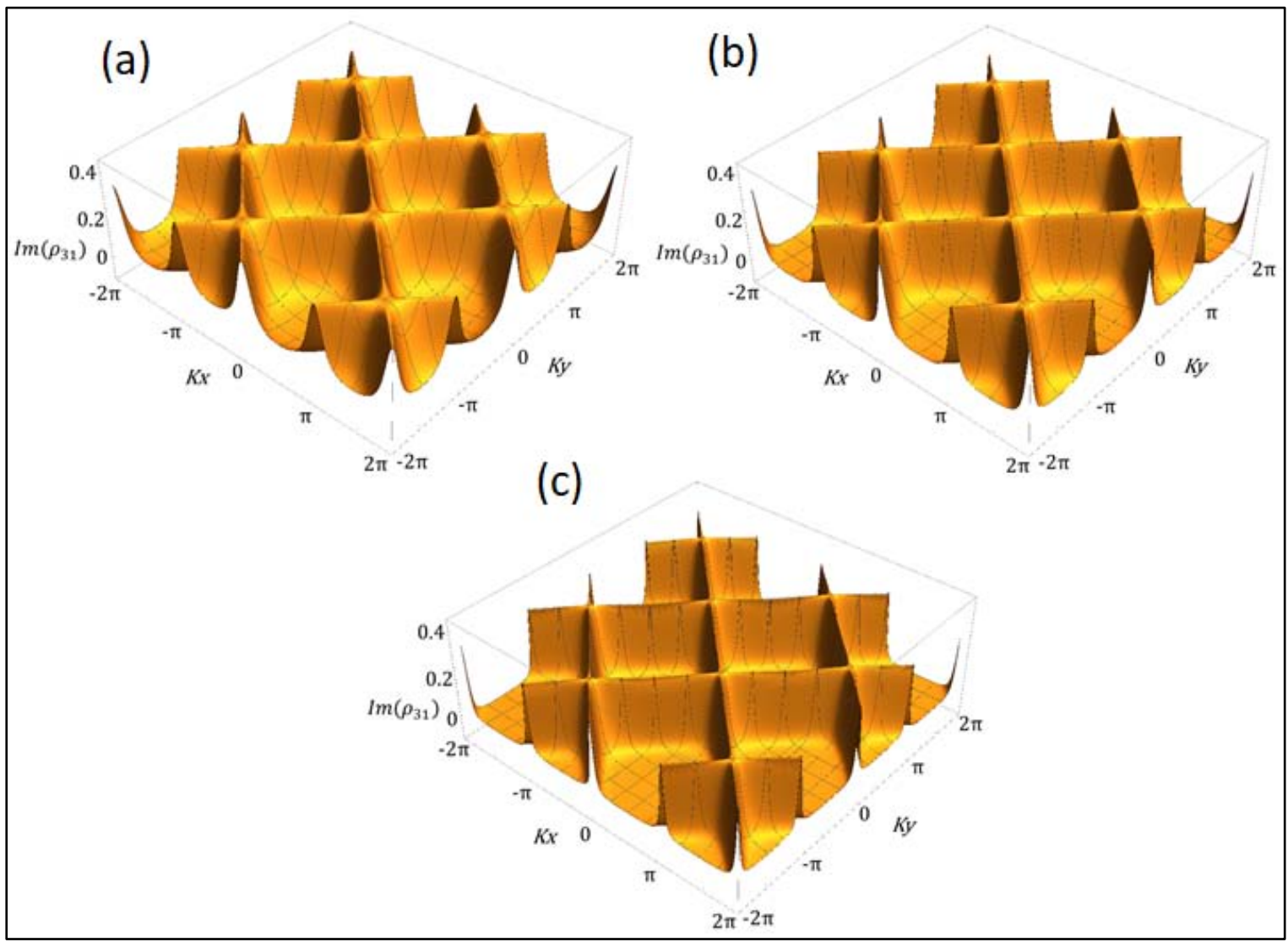

Fig. 7

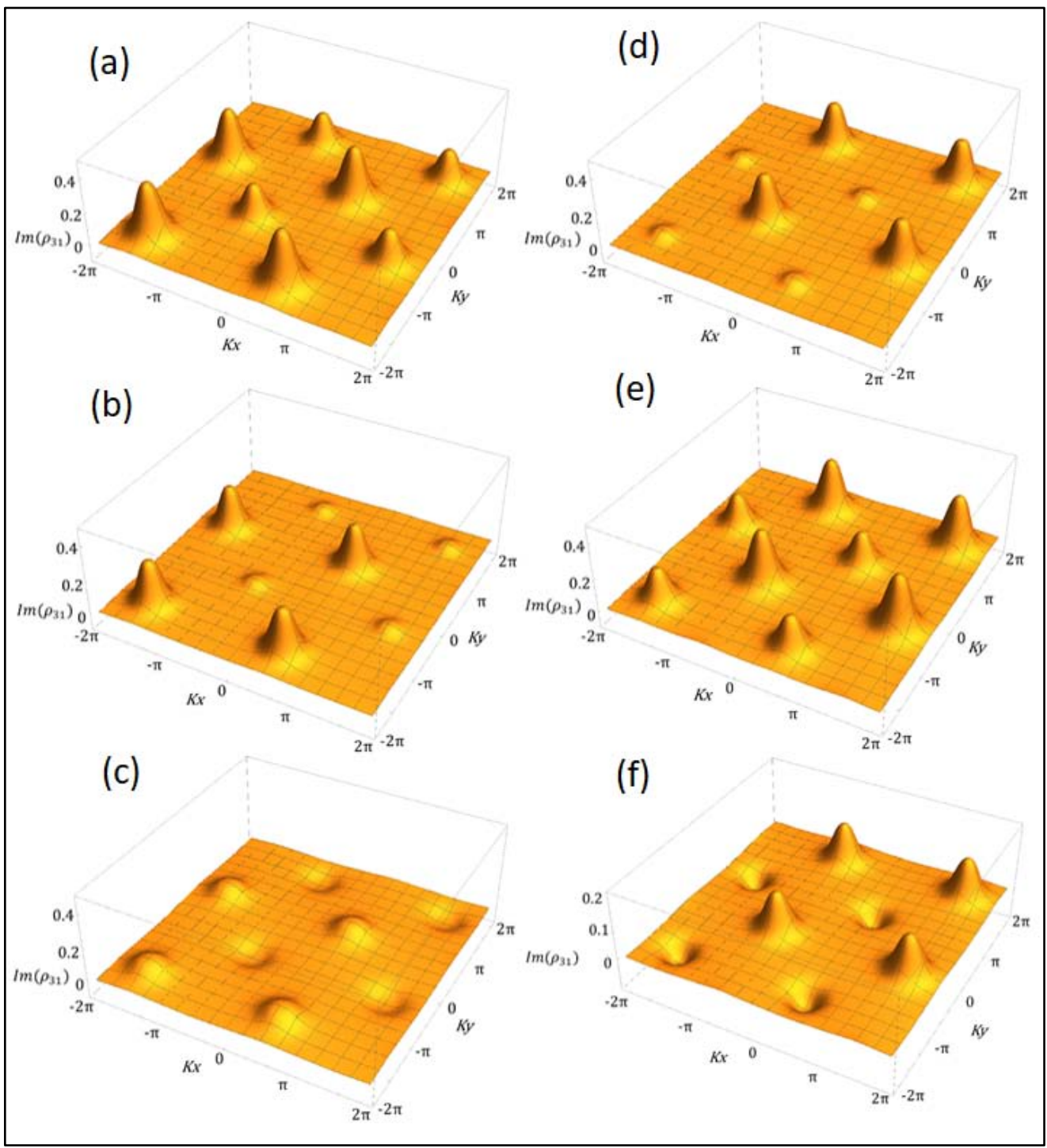

Fig. 8

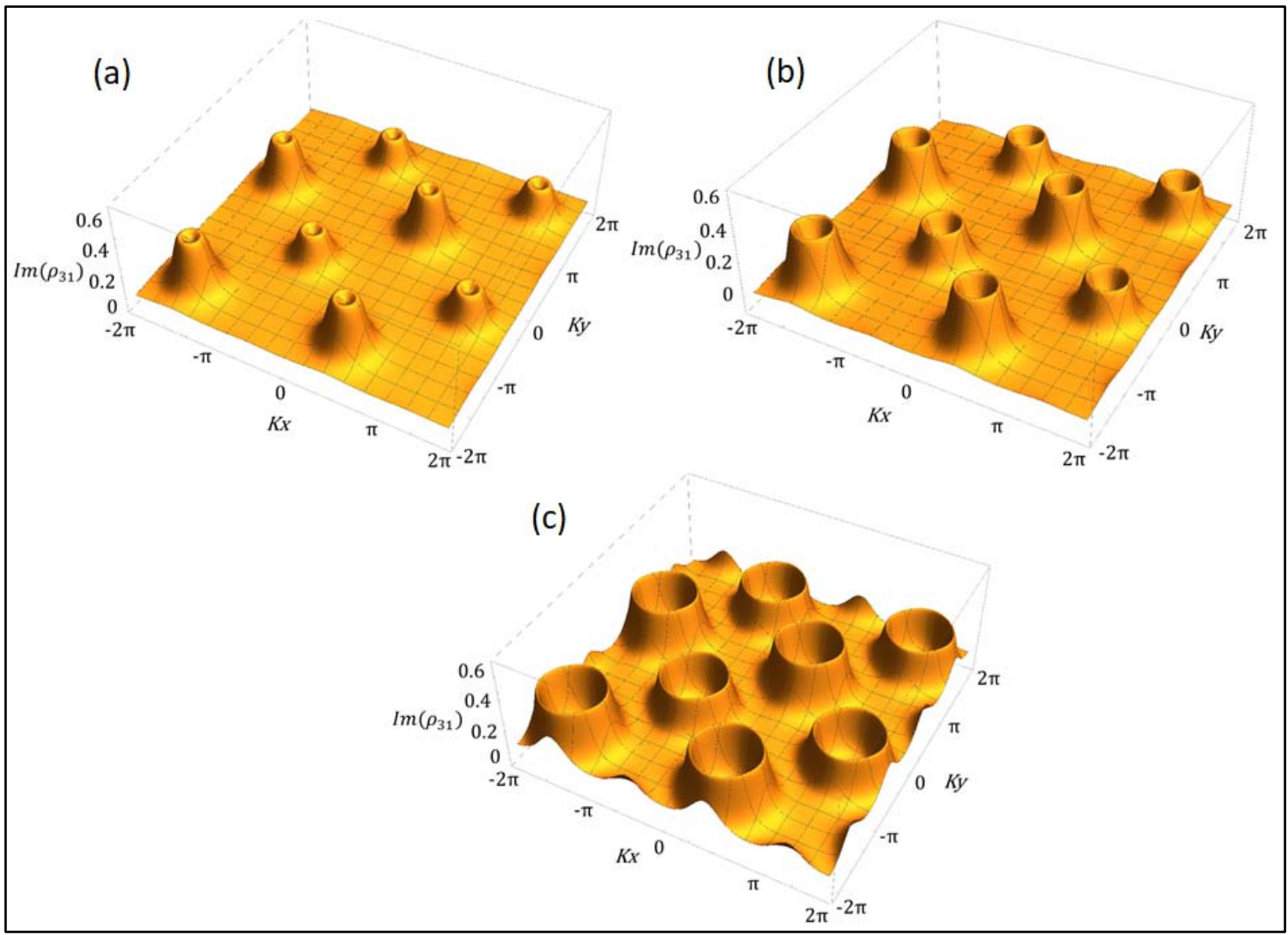

Fig. 9

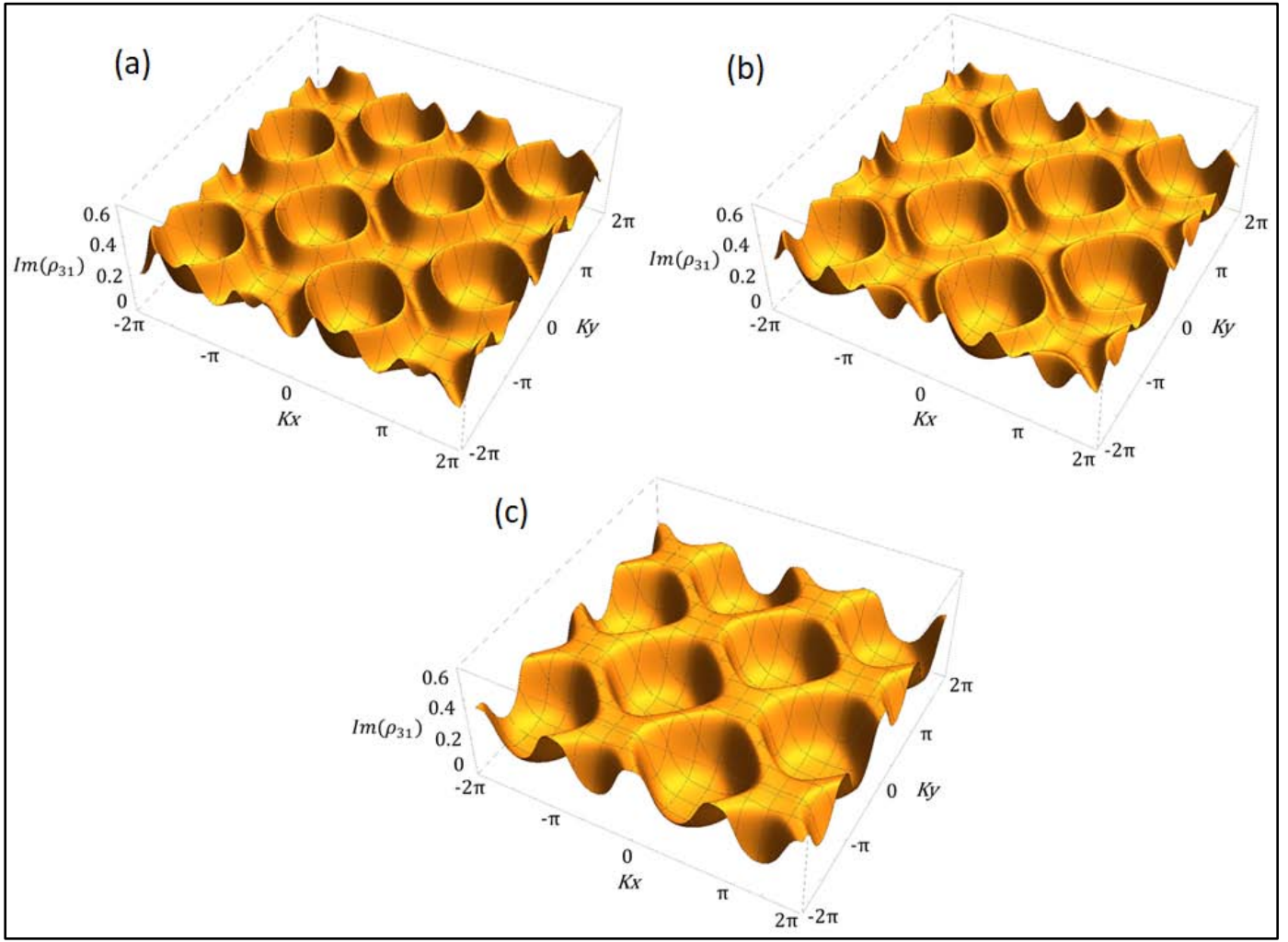

Fig. 10

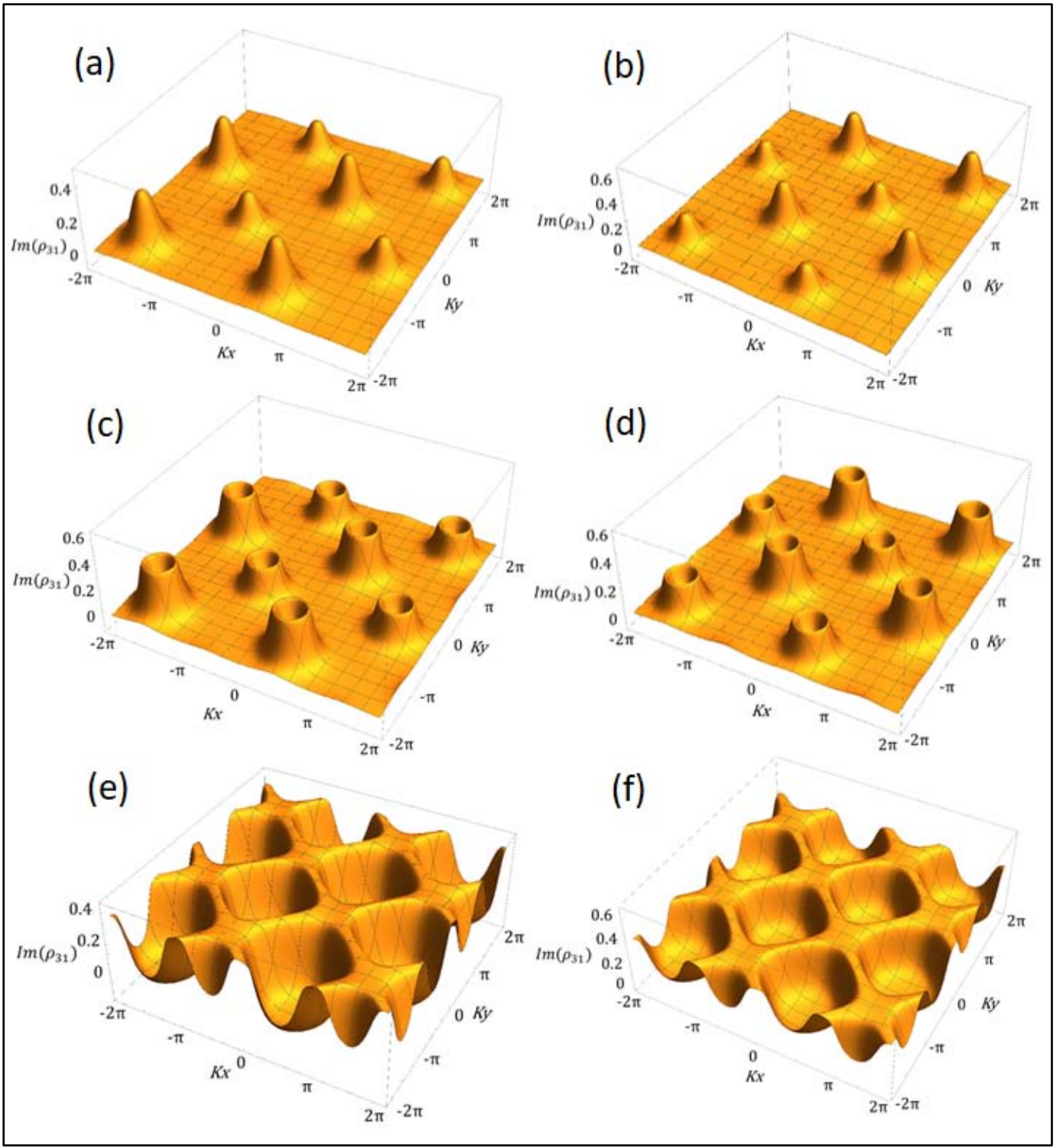

Fig. 11

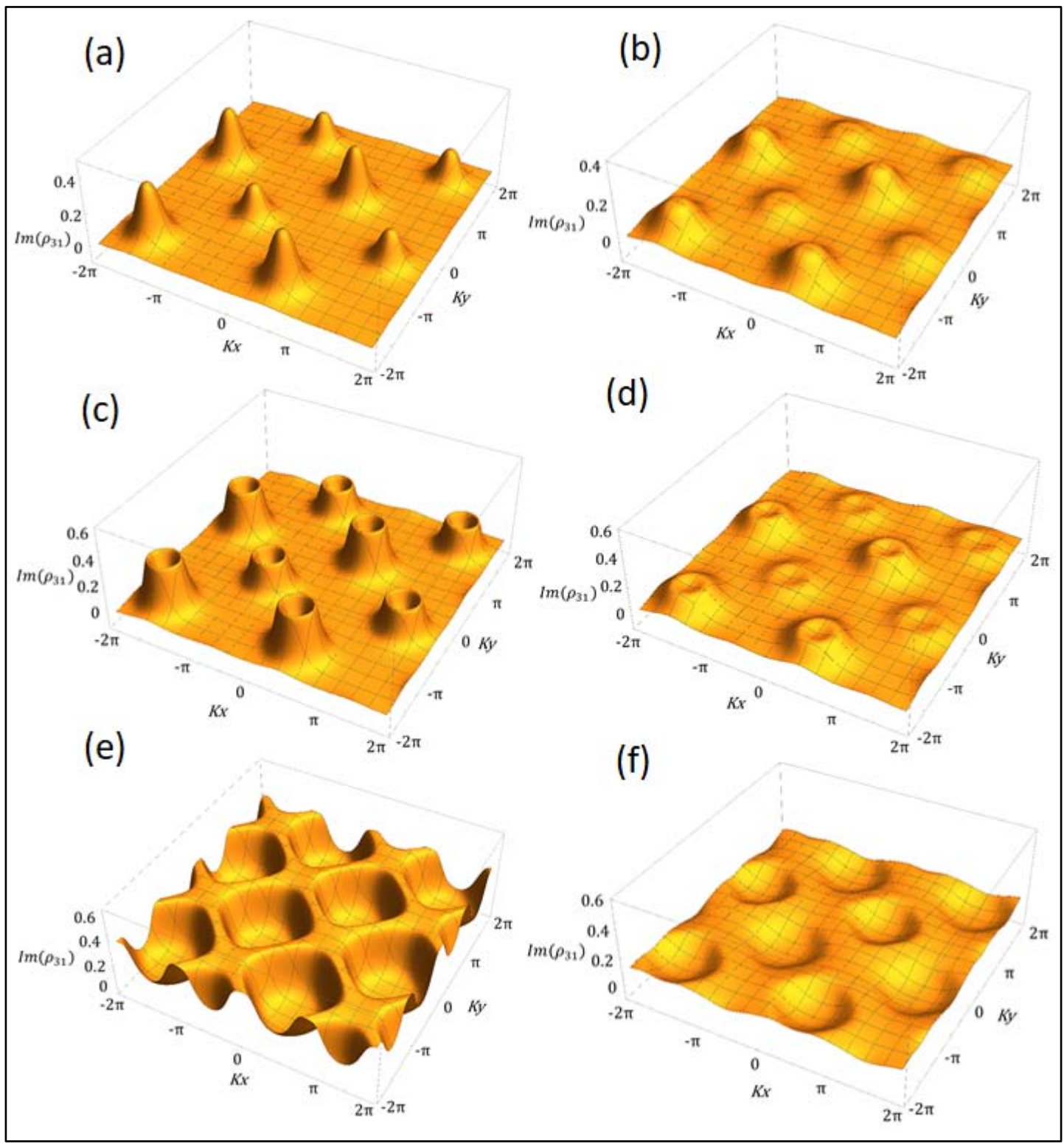

Fig. 12

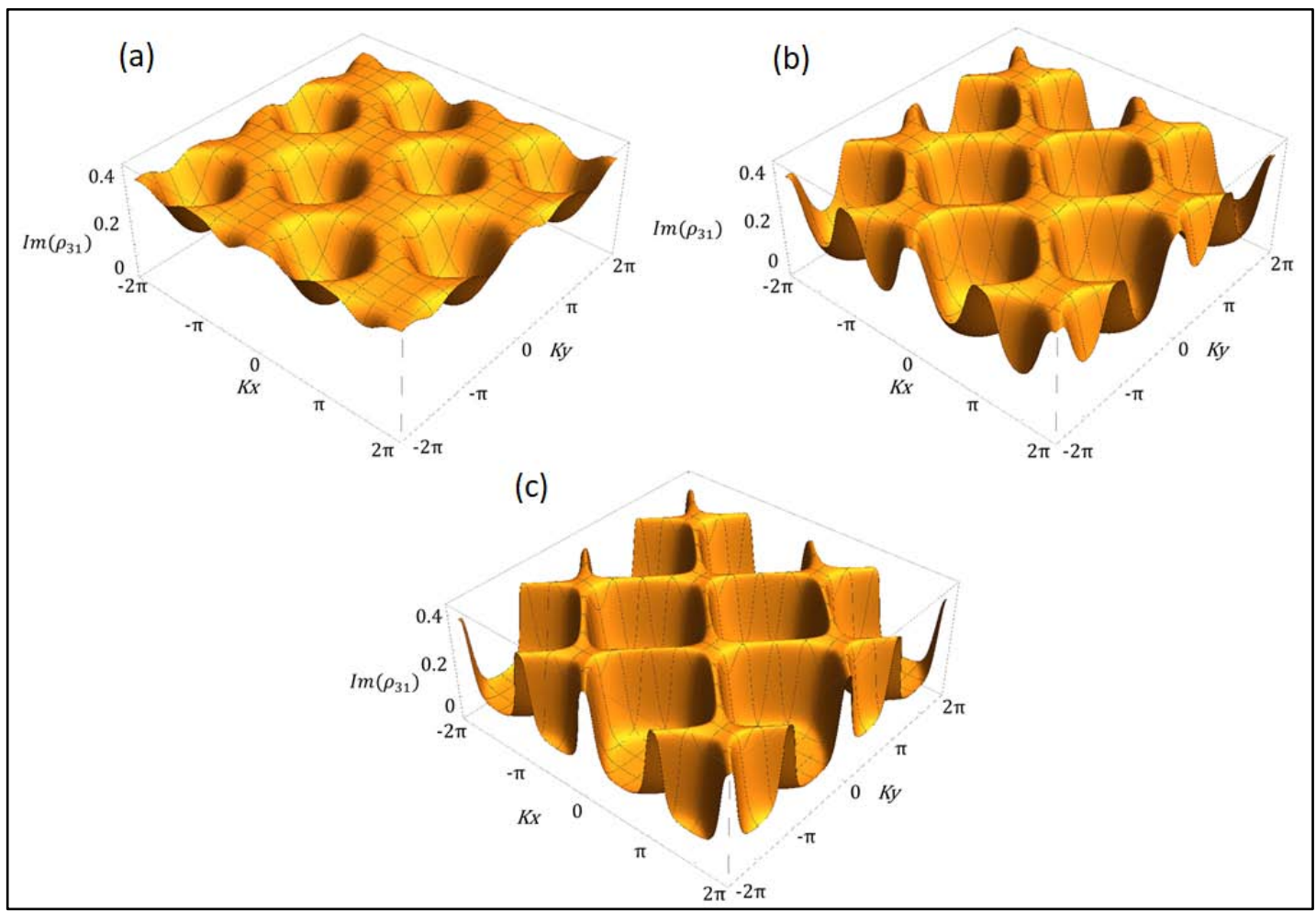

Fig. 13

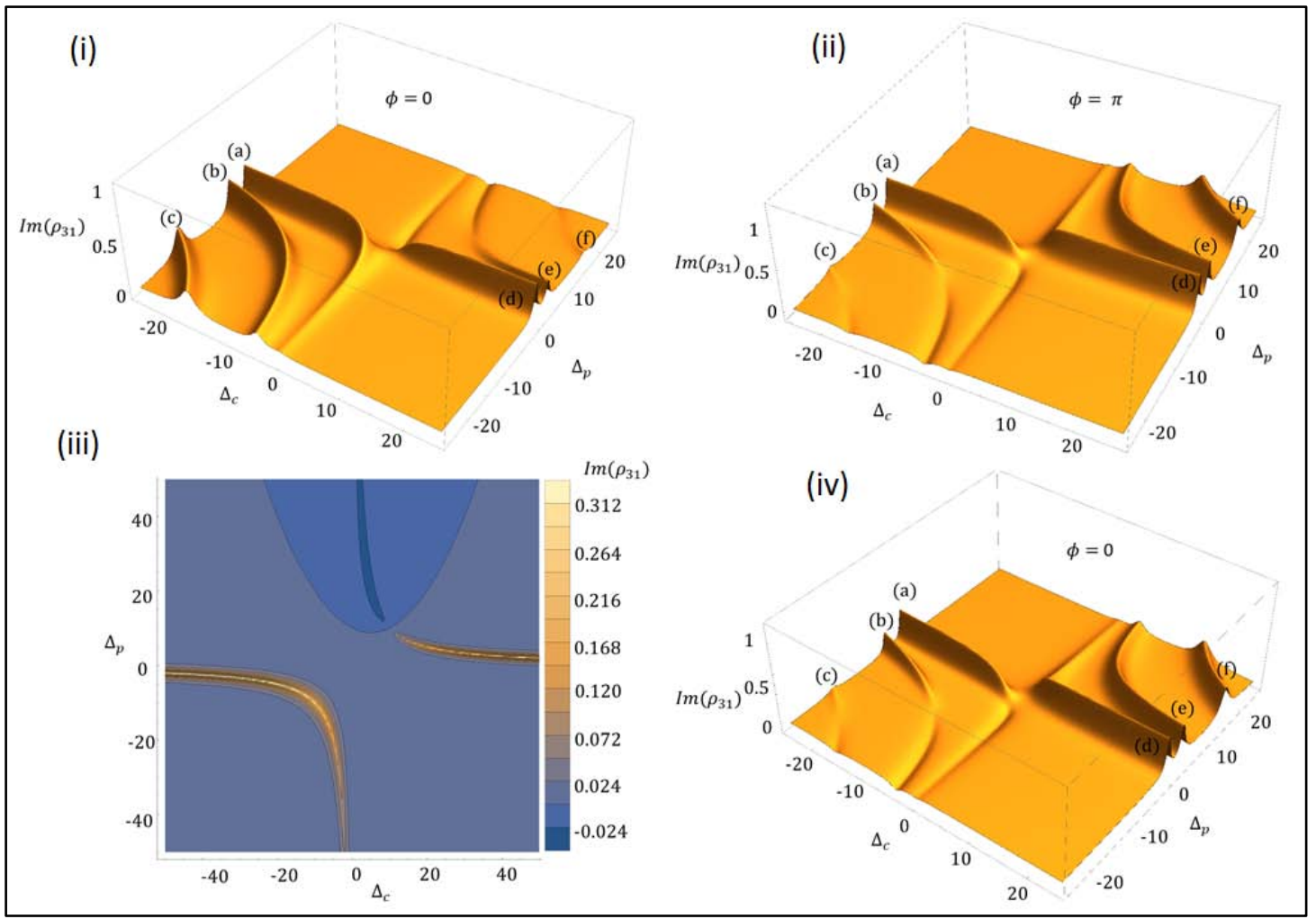

Fig. 14